\documentclass{iau}
\usepackage{amsmath}
\usepackage{graphicx}
\usepackage{multirow}
\usepackage{orcidlink}
\DeclareRobustCommand{\okina}{\raisebox{\dimexpr\fontcharht\font`f-\height}{\scalebox{0.8}{`}}}
\graphicspath{{./figures/}}
\usepackage{array}
\newcolumntype{P}[1]{>{\raggedright\arraybackslash}p{#1}}
\newcommand{\apj}{ApJ}
\newcommand{\apjl}{ApJL}     
\newcommand{\aap}{A\&A}
\newcommand{\solphys}{SoPh}
\newcommand{\ssr}{SSRv}
\newcommand{\nat}{Nature}

\begin{document}

\journaltitle{A Multi-Point view of the Sun: Advances in Solar Observations and in Space Weather Understanding}
\jnlPage{1}{12}
\jnlDoiYr{2025}
\doival{10.1017/xxxxx}
\volno{390}
\pubYr{2024}
\aopheadtitle{Proceedings IAU Symposium}
\editors{M. Romoli, L. Feng, M. Snow, eds.}

\title{\\ Intersecting frontiers for ground and space-based solar missions: symbiotic coordination between DKIST, PSP, and Solar Orbiter}
\lefttitle{Schad, Tritschler, and Cauzzi}
\righttitle{Intersecting frontiers for ground- and space-based solar missions}

\author{Thomas A. Schad$^1$\orcidlink{0000-0002-7451-9804}, Alexandra Tritschler$^2$\orcidlink{0000-0003-3147-8026}, and Gianna Cauzzi$^2$\orcidlink{0000-0002-6116-7301}}
\affiliation{$^1$National Solar Observatory, 22 \okina\={O}hi\okina a K\={u} Street, Pukalani, HI 96768, USA \\ $^2$National Solar Observatory, 3665 Discovery Drive, Boulder, CO 80303, USA}

\begin{abstract}
Three uniquely powerful solar and heliospheric facilities are now operational at the same time. The US National Science Foundation's Daniel K Inouye Solar Telescope, NASA's Parker Solar Probe, and ESA's Solar Orbiter each represent frontiers in space science, and each pursue richly tailored science missions. At the intersection of these missions, though, lie unparalleled opportunities for multi-vantage point science. This symbiotic relationship is especially pronounced during PSP's perihelia and Solar Orbiter remote science windows. As the most advanced solar polarimeter ever built, DKIST strengthens many of the multi-facility use cases by opening new diagnostic windows into solar magnetism---spanning the photosphere, chromosphere, and corona--- at unprecedented spatial, spectral, and temporal resolution. In this article, we report recent efforts to maximize the scientific potential of coordinated DKIST, PSP, and Solar Orbiter observations. Existing DKIST data from coordinated observations with Solar Orbiter and PSP are highlighted alongside some first investigations of these data. 
\end{abstract}


\maketitle

\section{DKIST, PSP, and Solar Orbiter: decades in the making} 

IAU Symposium \#390 assembled the global heliophysics science community at a most fascinating time, one in which three large science facilities have been brought online at nearly the same time. Together, the US National Science Foundation's Daniel K Inouye Solar Telescope \cite[DKIST:][]{rimmele2020}, NASA's Parker Solar Probe \cite[PSP:][]{fox2016}, and ESA's Solar Orbiter \cite[][]{Muller2020} offer an unprecedented multi-point view of the complex inner-heliosphere, extending from the solar interior to Earth. The realization of these independent projects only came about after tireless efforts to prioritize particular science objectives, develop the key technologies and methods to succeed, and then fund, plan, build, and execute. Particular milestones that helped prioritize these missions were the 2003 NRC Decadal Survey \citep{NRC2003} and the 2005 ESA's Cosmic Vision report \citep{ESA2005}; however, the initial dream of these missions predate these surveys, often by decades. 

PSP was launched in 2018, Solar Orbiter was launched in 2020, and DKIST finished its construction and moved to early operations in 2022. In this article, we highlight some of the science cases that unite these three missions while also discussing initial efforts to bring together all three of these assets during coordinated observations. These efforts offer great scientific promise but are not without practical challenges. Programmatically coordinating three independent missions operated under different observational models requires building a mutual understanding of mission strengths and operational constraints. As DKIST is currently in its Operations Commissioning Phase \citep[OCP,][]{Tritschler2022, Rimmele2024}, which implies elevated risks due to the novel capabilities and ground-based observing of this large facility, care is also required to optimize the potential for good science while accounting for these limiting factors.

\section{Science missions united through critical interfaces}

DKIST, PSP, and Solar Orbiter are motivated by a set of independent yet linked science objectives that are tied to long-existing, fundamental mysteries of the Sun and heliosphere, including (1) how is the upper atmosphere heated (including both the chromosphere and corona), and (2) how is the solar wind released and accelerated, and what creates the different types of wind. Despite significant progress in recent decades, these questions continue to evade full understanding. A very basic problem is that the energy dissipation is occurring at length scales in the solar atmosphere that are observationally inaccessible \cite[see, \textit{e.g.},][]{Bellan2020}; this implies that we must address these questions through intermediate and dynamically coupled diagnostic methods in combination with advanced numerical modeling. 

Guided by the above questions, key science objectives common to DKIST, PSP, and Solar Orbiter include: 
\begin{enumerate}
    \item To understand how solar magneto-convective energy is transported to the upper atmosphere, dissipated in the chromosphere and corona, and then further used to accelerate the solar wind.
    \item To connect magnetic structures and dynamics in the low atmosphere to magnetism and particles measured in situ.  
    \item To understand the physics that controls transient behavior on multiple scales, from pico-flares to active region eruptions.
\end{enumerate}
The frontiers that DKIST, PSP, and Solar Orbiter aim to advance require an understanding of the energy flow across key boundaries. From the perspective of the lower solar atmosphere, while we still do not have complete observational knowledge of photospheric magneto-convection, radiative MHD simulations capture many of the essential elements, leading to a high level of insight into the bottom boundary of the heliosphere \citep{Nordlund2009, Rempel2014}. Meanwhile, from the heliospheric vantage point at 1 AU, both in and out of the ecliptic, we have inventoried a diversity of solar wind types classified, in part, through speed, composition, and potential solar and interplanetary sources \citep{Verscharen2019}.

Issues arise as we try to solidify the connections between these two vantage points, especially given the sparse sampling of in-situ data. We have many tools at our disposal, including measuring and extrapolating magnetic fields, mapping elemental compositions, tracing transient activity, and so forth. However, we know that these methods often have limits. Moving upward into the solar atmosphere from the photosphere, we encounter the problematic interface where $\beta \approx 1$ as the plasma transitions from being pressure-dominated to magnetic field-dominated \citep{Wedemeyer2009, Carlsson2019}. Here, within the chromosphere, there is a plethora of potential processes at work that are very difficult to observationally constrain, especially as the plasma becomes partially ionized, radiative line diagnostics are formed in non-thermodynamic equilibrium, magnetic fields are weaker and more difficult to measure, and wave and mass motions occur on fast timescales and within fine structures. Tracing the energy transport from the photosphere to the upper atmosphere, and then understanding how it is dissipated, is consequently very challenging. 

Moving towards the Sun from the 1 AU perspective, there are similar challenges, as the solar wind properties are modified in transit from the inner heliosphere, especially as the plasma $\beta$ here is generally larger than one such that the field is dragged by the wind \citep{viall2020}. Consequently, corotating interacting regions lead to rarefaction and compression of the plasma, leading as well to interplanetary shocks. Wave interactions occur during the transit of the wind through the heliosphere, and the imprints of the accelerating mechanisms at work near the Sun get blurred and washed out.  

By uniting the assets of DKIST, PSP, and Solar Orbiter, we can attempt to form a bridge between these disparate vantage points. The frontier of observational solar physics is to better probe the chromospheric to corona mass and energy cycle, where the magnetic field is dominant but poorly measured. The frontier of space-based missions lies in probing the inner heliosphere directly in situ. Together, we can then hopefully build a unified view of energy transport from the solar interior to the inner heliosphere.

\section{Tracing energy to the base of the solar atmosphere}

\begin{figure}
    \centering
    \includegraphics[width=0.485\linewidth]{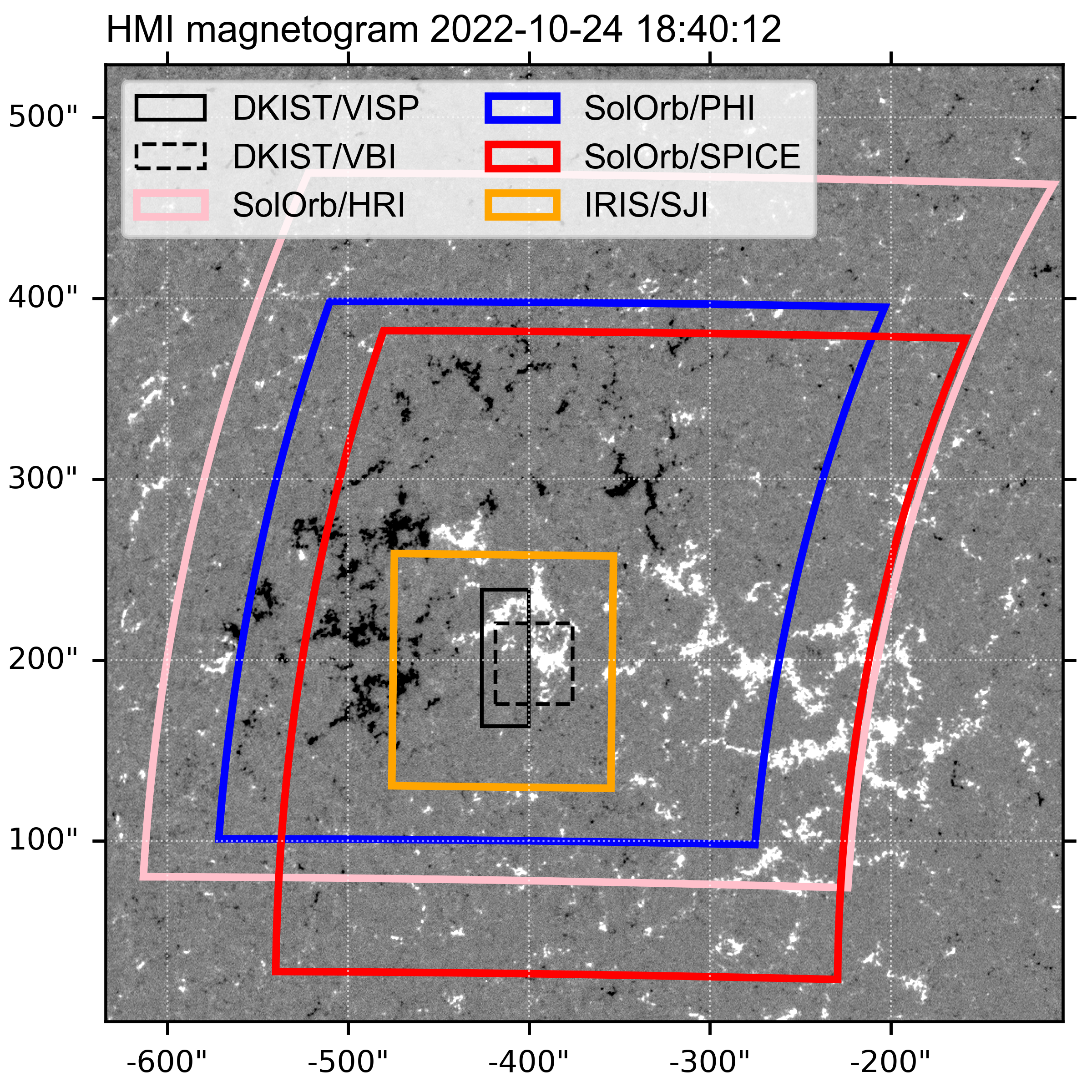} 
    \includegraphics[width=0.485\linewidth]{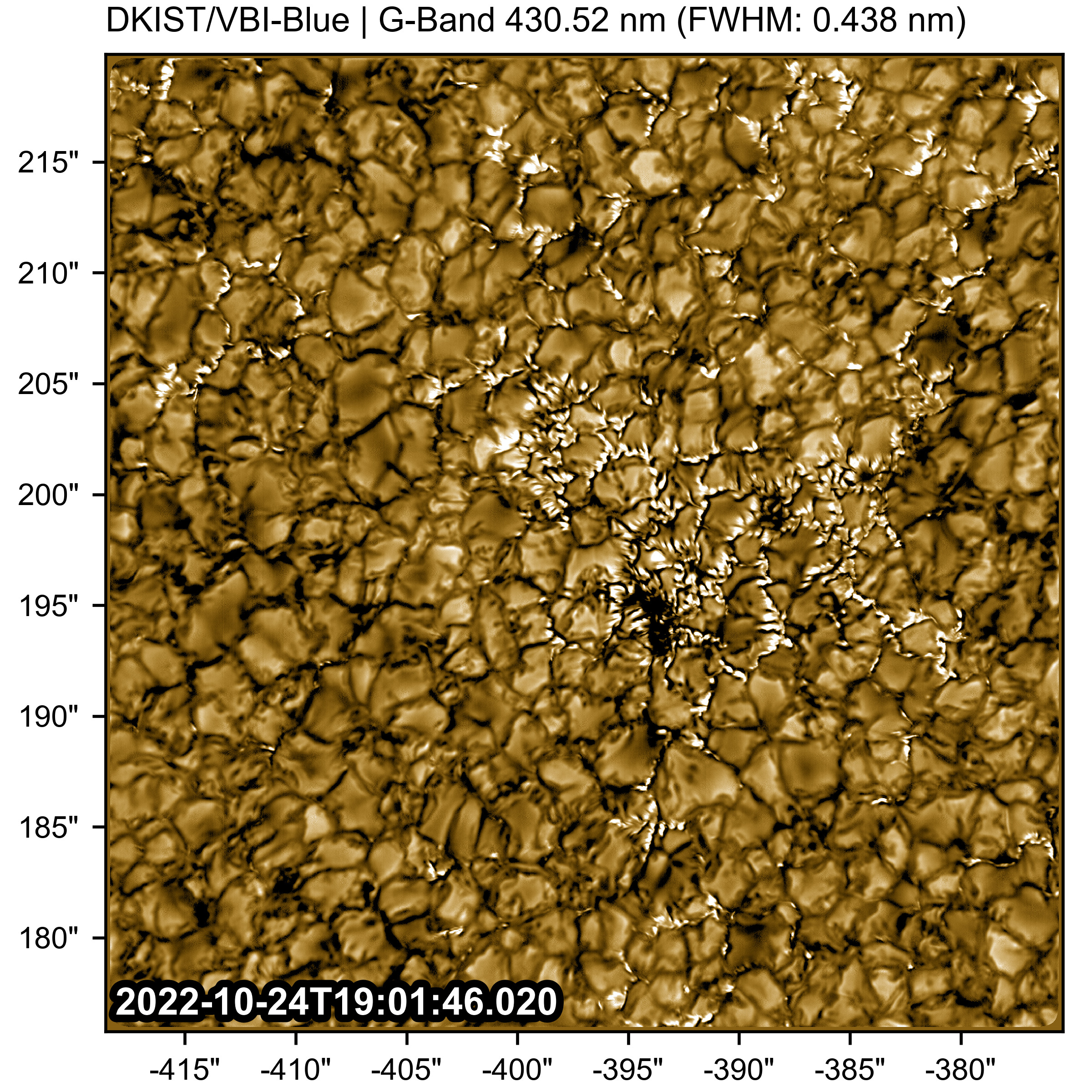}\\
    \includegraphics[width=0.485\linewidth]{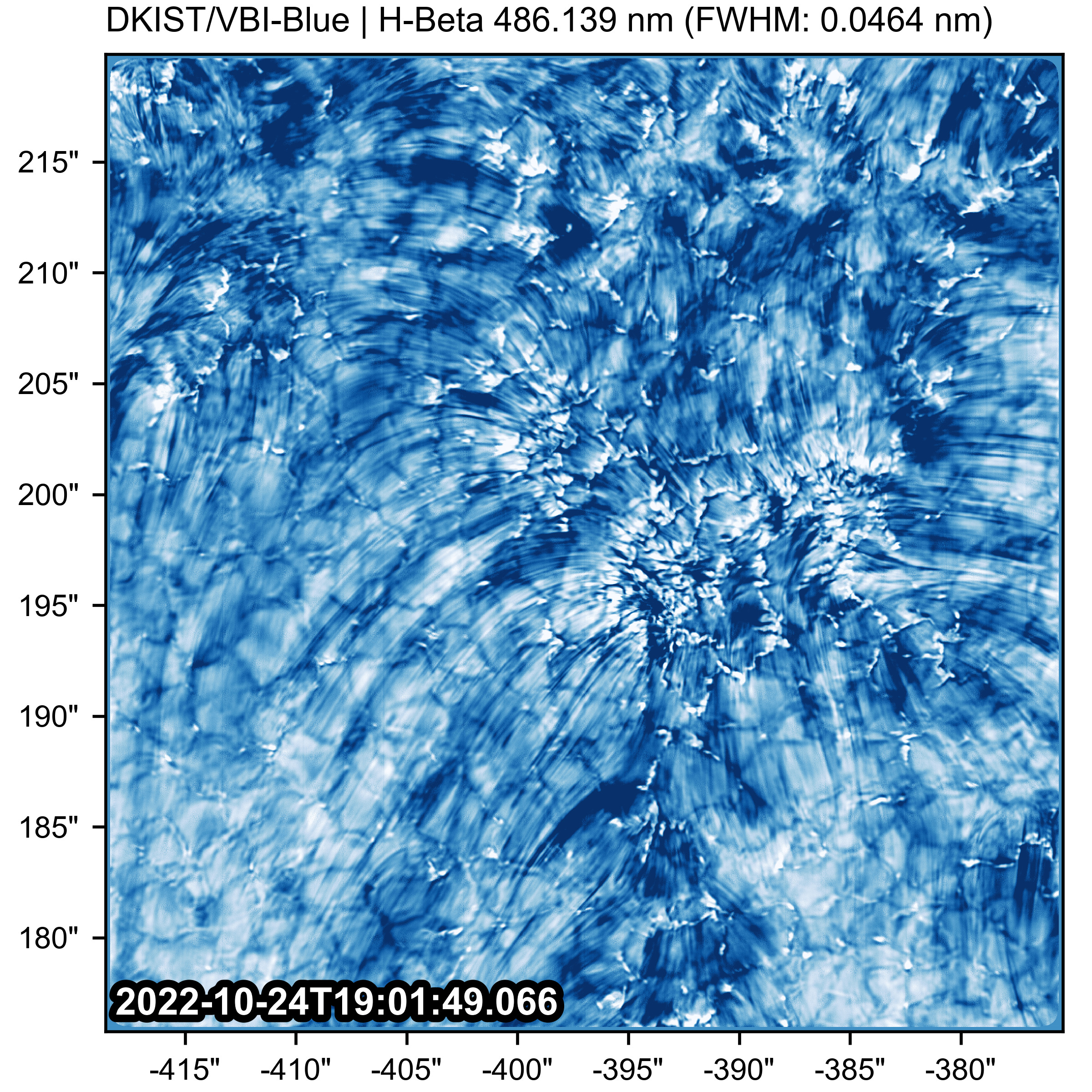}
    \includegraphics[width=0.485\linewidth]{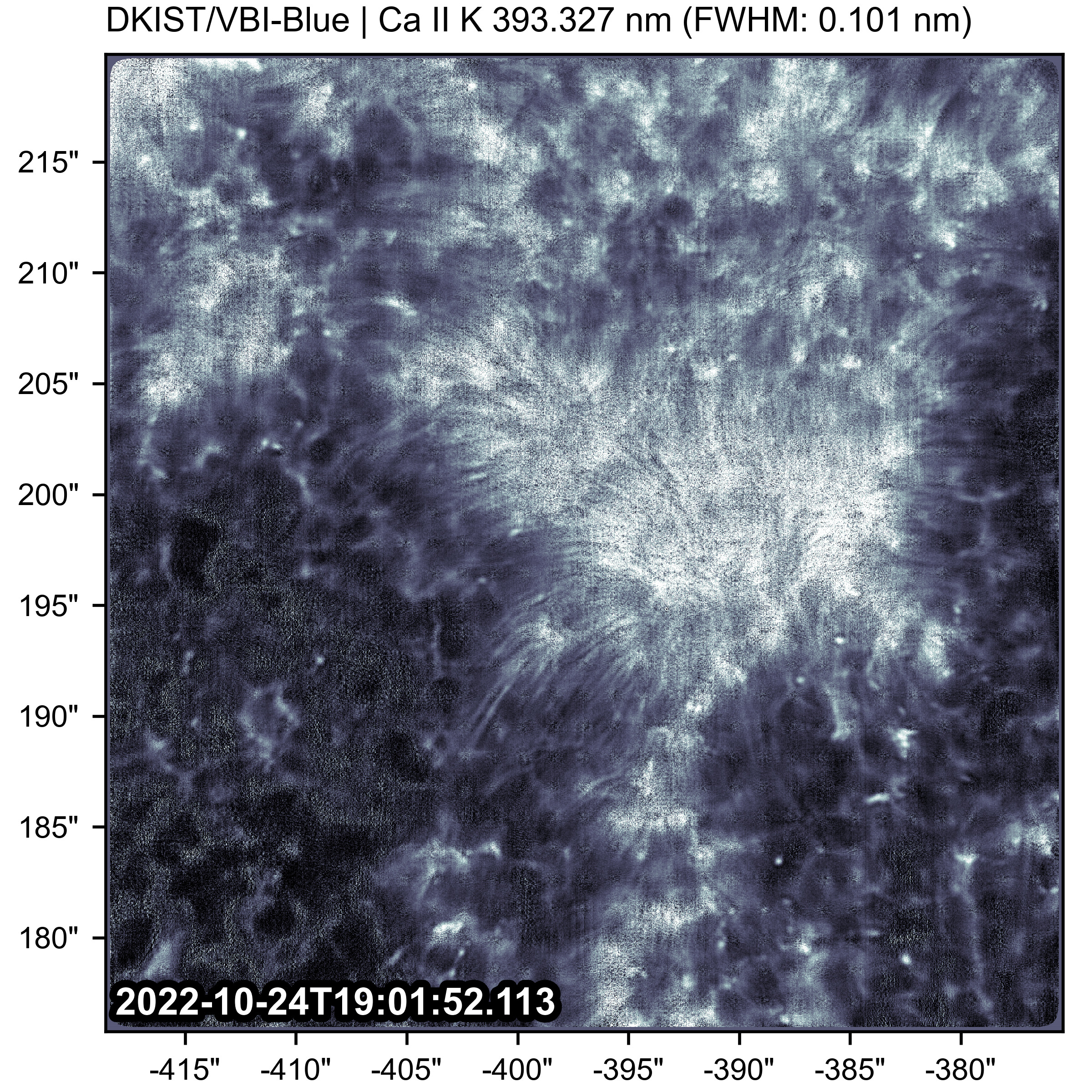}
    \caption{Examples of high spatial resolution DKIST/VBI data acquired on 24 Oct 2022 during Solar Orbiter's RSW \#5. These data are from DKIST Experiment EID\_1\_112. For more details of coordinated observations during this window see \citet{barczynski2025}.}
    \label{fig:vbi_2022}
\end{figure}

The advent of near Sun in-situ data, starting with PSP's first perihelion in 2018, has provided key insights into the energy source driving the solar wind and how it may be traced to the lower atmosphere. As \cite{Raouafi2023} states, PSP has revealed the highly Alfv\'enic character of the near Sun solar wind, pointing to a magnetic origin at the Sun, which these authors argue may find its roots in the small-scale reconnection of opposite-polarity flux that forms so-called jetlets. Moreover, this small-scale activity could be linked to the observed magnetic switchbacks in the near Sun solar wind \citep{Bale2019, Kasper2019}. It's clear, however, that to unambiguously form these connections requires tracing both the energy flow and magnetic field upward from the photosphere and backward from the solar wind during perihelion events, especially near or within the Alfv\'en surface where dynamical connections may remain coherent. 

Resolving the energy content of the lower atmosphere is one motivation behind DKIST's large four-meter aperture and its high-resolution capabilities. Using MURAM 3D radiative MHD simulations of the small-scale solar dynamo, \cite{Rempel2014} estimates that 50\% of the magnetic energy distribution in the solar photosphere occurs on lengths scales inaccessible from meter-class telescopes. Thus, with the previous generation of telescopes, we are essentially blind to about half of the energy available to power the upper atmosphere; meanwhile, approximately 90\% is expected to be observable by DKIST at the diffraction limit achievable using its integrated high-order adaptive optics system. 

By using speckle reconstruction techniques, the DKIST Visible Broadband Imager (VBI) delivers broadband images in multiple photospheric and chromospheric diagnostics with a cadence of approximately 3 seconds and spatial resolution of up to 11 and 16 milliarcseconds in its blue and red channels \citep[$\sim$ 15 and 23 km at the solar surface, respectively,][] {Woeger2022}. Ongoing projects include the study of striated features within inter-granular lanes that provide insights into the magnetic field distribution within the network and inter-network regions at the diffraction limit.

In addition to high spatial resolution, DKIST's large effective aperture and rapid imaging cameras provide new insights into how the lower atmosphere can inject energy into the upper atmosphere. Results from \cite{Fischer2023} have demonstrated how shock waves propagating horizontally away from G-band bright points, most likely a signature of localized heating, can be resolved in chromospheric imaging with the Ca II K VBI filter. Furthermore, the bright points driving these events exhibit vortical twisting behavior. Although such small-scale dynamics in the lower atmosphere may play a sizable role in the production and acceleration of the solar wind, it remains hard to unambiguously quantify the energy injection rates into the solar wind. Using the advanced simulations of the Bifrost code, \cite{Finley2022} have shown that twisted flux tubes may launch torsional Alfv\'en waves up through magnetic funnels and subsequently enhance the turbulent generation of magnetic switchbacks in the solar wind. Meanwhile, Solar Orbiter observations of coronal holes suggest that magnetic reconnection leading to pico-flares may be frequent enough to power the solar wind \citet{Chitta2023}.

In Figure~\ref{fig:vbi_2022}, we show a few examples of VBI images acquired in October 2022 during the Solar Orbiter Remote Science Window \#5. In this case, the science objectives included tracing fine-scale flow fields around an active region to understand its long-term evolution, as can be studied over multiple weeks from Solar Orbiter's orbit. For more information about these observations, please see the forthcoming article by \cite{barczynski2025}. 

\section{Solar magnetism through the upper solar atmosphere}

\begin{figure}
    \centering
    \includegraphics[width=0.95\linewidth]{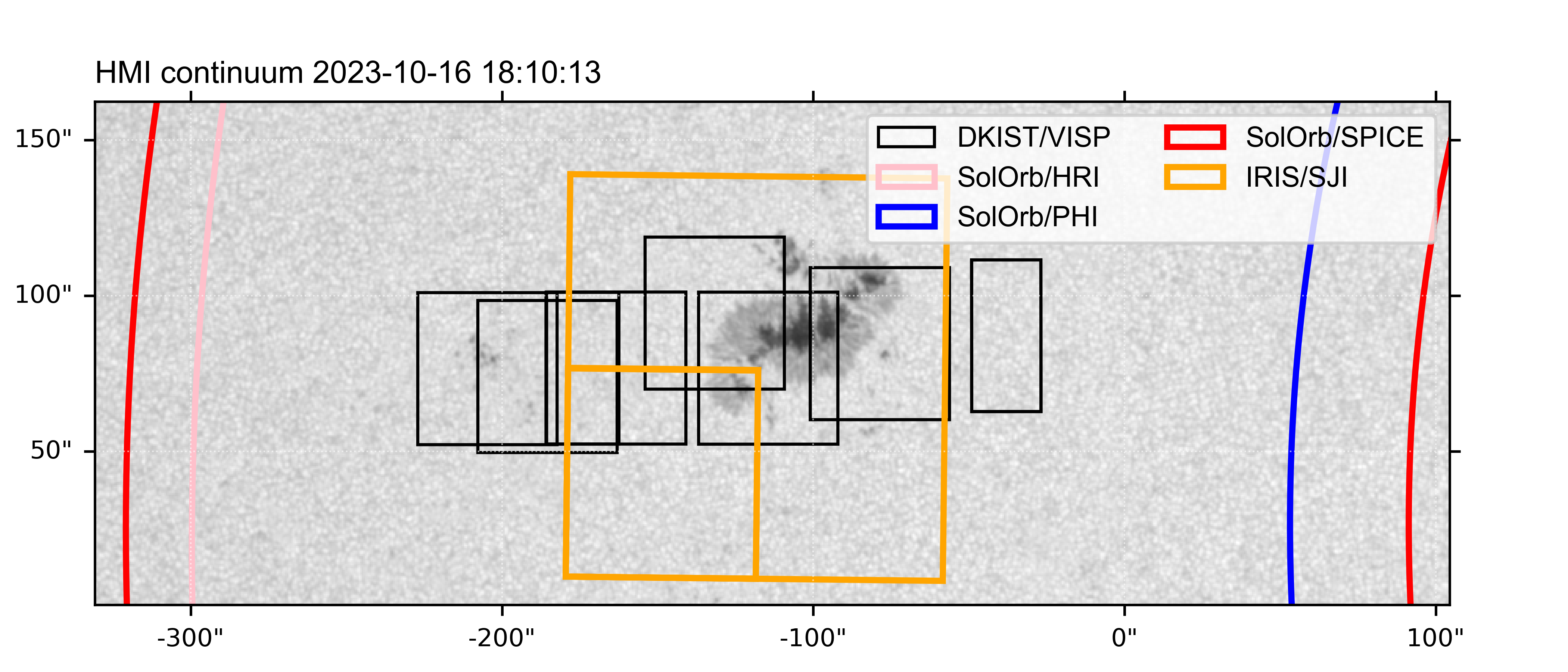} \\ 
    \includegraphics[width=0.95\linewidth]{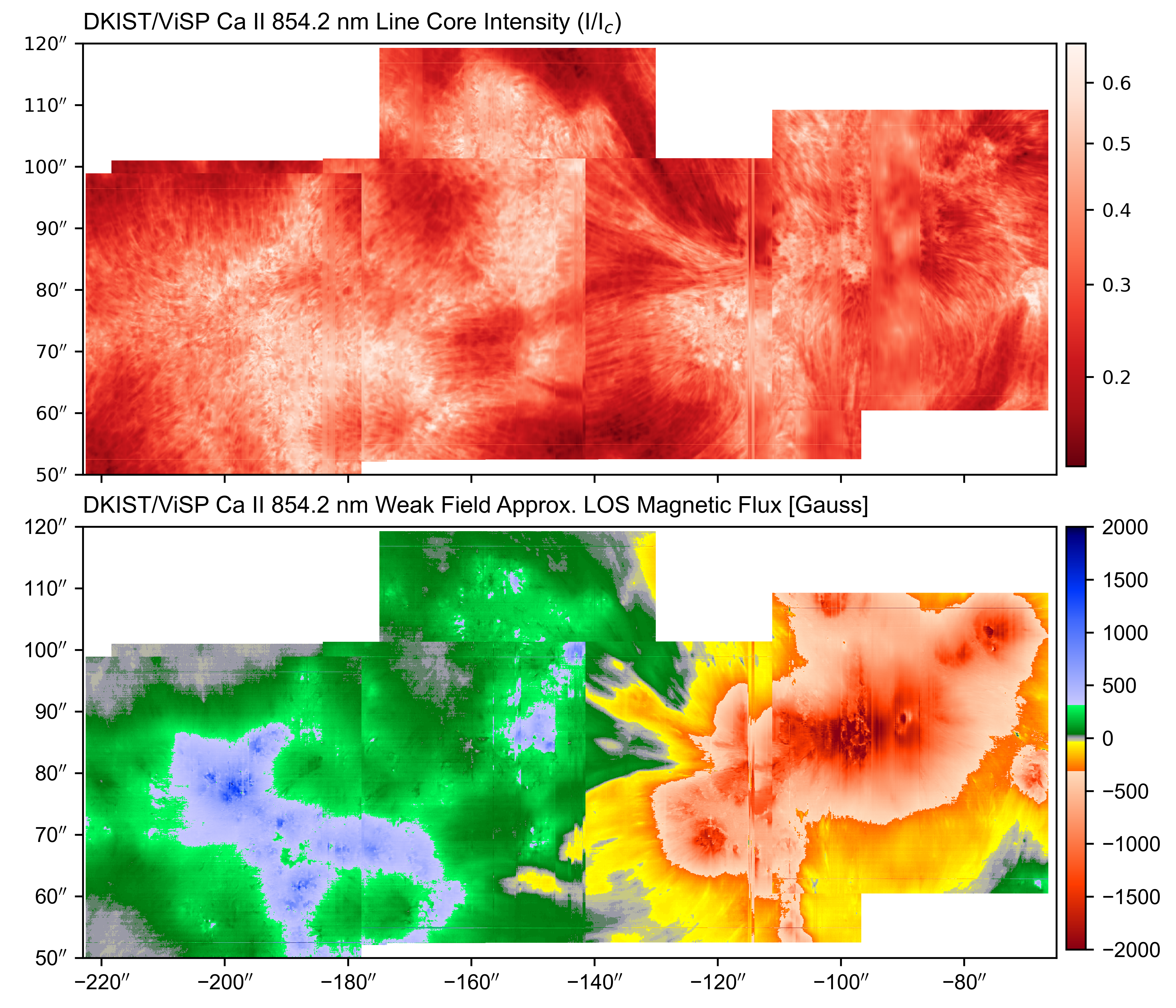}
    \caption{Example chromospheric polarimetric data acquired by DKIST/ViSP on 16 October 2023 during Solar Orbiter RSW \#11. The top panel shows select FOVs of the DKIST/ViSP Arm \#3 data alongside Solar Orbiter and IRIS data. Not all available data are shown, and the analysis is preliminary (for display purposes only). The bottom panel results from the weak field approximation applied to the Ca II line core ($\pm0.25\AA$) using Equation 7 of \cite{MartinezGonzalez2009}.} 
    \label{fig:visp_2023}
\end{figure}

The role of the magnetic field in channeling convective energy into the atmosphere is widely accepted, and small-scale opposite polarity fields have been invoked as a critical element in magnetic-reconnection driven solar wind, as well as active region coronal loop heating \citep{Chitta2017, Chitta2023, Raouafi2023}.   That said,  connecting the photospheric magnetic field with the chromosphere and above, especially on small scales, remains very difficult, given that the non-force-free nature of the photosphere limits our ability to extrapolate the field to larger heights \citep{deRosa2009, Peter2015}.  Thus, high-resolution polarimetry of the chromosphere and transition region becomes essential to discern the true influence of magnetic fields above the photosphere, including the build-up of magnetic energy on large scales and the small-scale driving of the solar wind.  

Diagnostics of the magnetic field in the upper atmosphere require advanced methods not only to measure very weak signals but also to interpret spectral signatures formed outside of local thermodynamic equilibrium. The biggest challenge is that such studies need to be carried out with sufficient temporal resolution and polarimetric sensitivity to perform full Stokes spectral inversions in these highly structured and dynamic environments. This is very difficult, if not unreachable, without a large aperture telescope like DKIST. That said, \cite{gosic2018} using the Swedish Solar Telescope found some initial evidence that network emerging flux is able to reach the chromosphere.

DKIST has multiple spectropolarimeters that can probe the magnetic field from the photosphere to the chromosphere and even in the off-limb corona. In combination with Solar Orbiter, DKIST polarimetry can further be used for multi-vantage point magnetic field mapping. One advantage of such observations is in the disambiguation of photospheric magnetic fields \citep{Valori2022}; however, the true strength of DKIST lies in its multi-wavelength polarimetric measurements.   The Visible Spectropolarimeter \citep[VISP:][]{deWijn2022}, for example, can observe simultaneously in three different wavelength channels between 380 and 900 nm, providing multiple diagnostics to reliably constrain spectral inversions over a large span of the atmosphere.

\cite{daSilvaSantos2023} published some of the first high-sensitivity maps of the total linear and circular polarization in the chromospheric Ca II 854 nm line, using data obtained during the disk passage of PSP just after its perihelion \#12. This was a first step prior to more routine full spectral inversions of these types of diagnostics, which are being pursued by the National Solar Observatory (NSO) Community Science Program for Level 2 data products. 

In Figure~\ref{fig:visp_2023}, we show some additional chromospheric polarimetry data obtained on 16 October 2023 during Solar Orbiter RSW \#11. To increase the field of view coverage, multiple pointings of the telescope were used, which also coincided with different lock points for the DKIST adaptive optics system. Co-temporal Solar Orbiter data, including with the Polarimetric and Helioseismic Imager (PHI)  \citep[SO/PHI:][]{Solanki2020}, will be very useful to study the use of multi-vantage point and multi-height polarimetry of active regions. 

Of course, mapping the magnetic field in the chromosphere is only part of the problem. For the question of energy transport, an even more challenging problem is connecting the magnetic field and dynamical motions in the chromosphere to discern candidate energy dissipation mechanisms. Investigating observationally the relationship between radiative/kinetic losses and current densities and fluid motions is one important approach. \cite{Anan2021}, for example, reviewed potential heating mechanisms in plage/network as they relate to properties of the local magnetic field configuration. However, even with a four-meter aperture telescope like DKIST, the combination of fast timescales (on orders of 10s of seconds or less), the need to cover spatial structures simultaneously, and requirements for spectropolarimetric coverage and sensitivity motivates the development of new observing methods. To this end, the DKIST Diffraction-Limited Near-Infrared Spectropolarimeter (DL-NIRSP) is advancing integral field spectropolarimetry for simultaneous spatial and spectral coverage over 2D fields-of-view at rapid cadence \citep{Jaeggli2022}. Recently, the DL-NIRSP’s fiber-based integral field unit was replaced by a cutting-edge nano-machined image slicer device manufactured by Canon Incorporated \citep{lin2022,anan2024}. This upgrade has increased the instrument's spectral throughput while also greatly enhancing flat-field stability and polarization accuracy. While calibration work remains, DL-NIRSP will certainly provide new methods for probing magnetic fields and dissipation in the lower atmosphere and in the chromosphere. 

\section{Unveiling coronal structure and dynamics}

It is usually accepted that the energy responsible for heating the low solar corona can be both in the form of Alfv\'enic waves, as observed e.g. by the HAO/CoMP instrument \citep{tomczyk2007}, as well as currents and available free energy stored within the low $\beta$ plasma.  \cite{Hahn2013} found some evidence based on Hinode/EIS line widths that these waves are dissipated in the low corona, although these observations can be difficult to interpret and calibrate. With its coronagraphic capabilities and large aperture, DKIST offers the potential to much better resolve, both spatially and temporarily, the spectrum of velocity oscillations present in the low corona and up to about 1.5 solar radii. This is done by observing the off-limb forbidden emission lines, like the Fe XIII 1074 and 1079 nm coronal lines. Better resolution of the wave spectra is important not only for heating the lower corona but also for connecting the low corona to the inner heliosphere and the magnetic field fluctuations observed therein \cite{Cranmer2017}.

\begin{figure}
    \centering
    \includegraphics[width=0.99\linewidth,trim={0 5.1cm 0 0},clip]{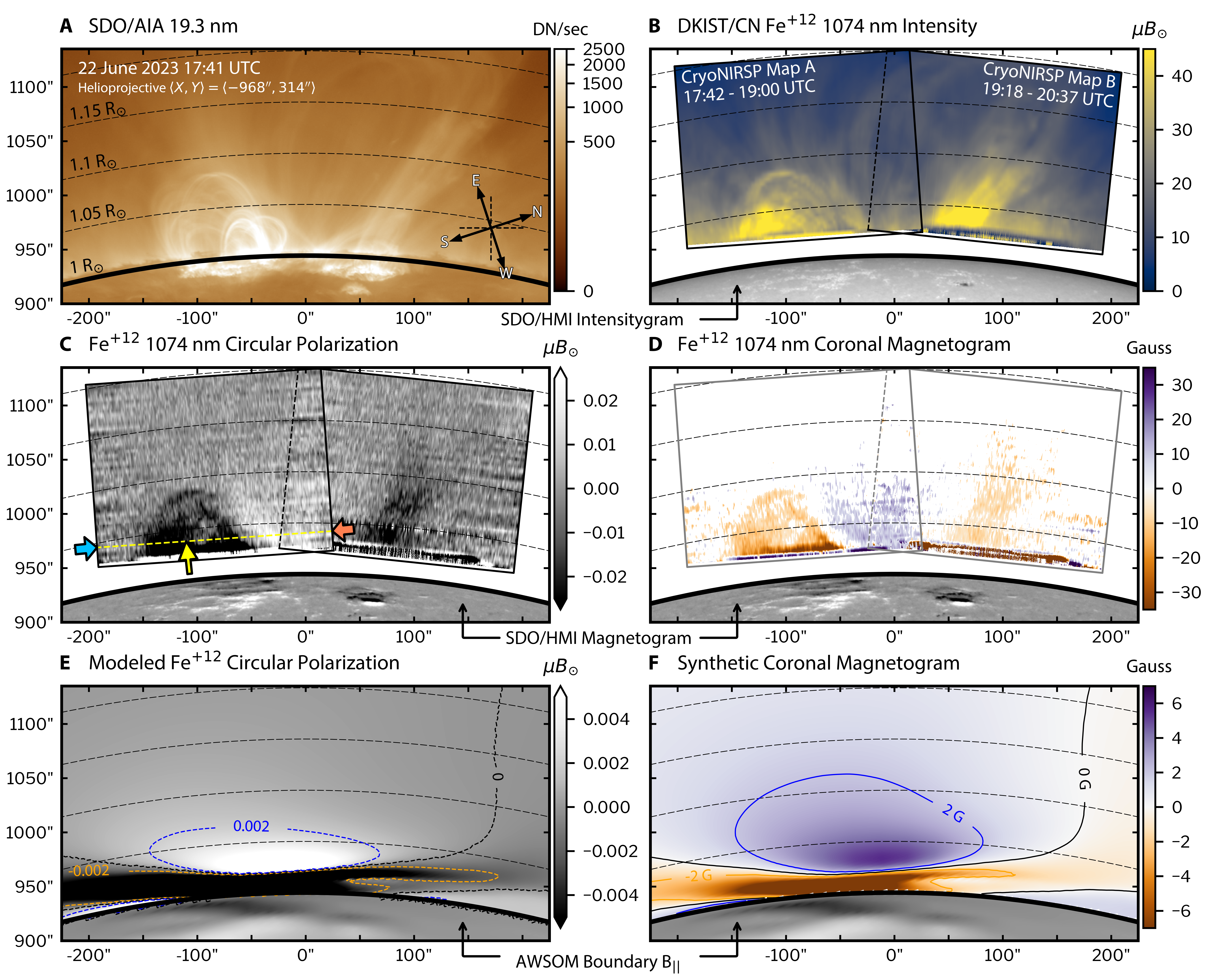}
    \caption{First DKIST Coronal Magnetograms as first reported in \cite{schad2024_science}. a) SDO/AIA 193\AA\ image cropped and rotated to the geometry of DKIST/Cryo-NIRSP observations acquired on 22 June 2023. This target was selected on the basis of the consensus predictions of the solar source regions connected to the Parker Solar Probe during its 16th perihelion. Vertical coordinates give arc seconds from the center of the solar disk. (B) the peak line amplitude of Fe \textsc{xiii} 1074 nm observed within the overlapping raster maps of CryoNIRSP in units of parts per million of the solar disk intensity, i.e., $\mu$B$_{\odot}$. (C) Peak red-wing amplitude of the measured antisymmetric circular polarized Fe XIII 1074 nm profile. (D) Inferred coronal longitudinal magnetogram in units of Gauss as inferred from the weak-field approximation fitted to the circularly polarized profiles.}
    \label{fig:stokesV}
\end{figure}

The advantages of the forbidden coronal emission lines also extend to polarimetric diagnostics of the coronal magnetic field. Indeed, measuring the strength of the coronal magnetic field from the ground is one of the key science objectives for DKIST, and its off-axis internal occulted coronagraphic design advances this goal by using the Zeeman effect.   The coronal Zeeman effect signal is very weak; relative to the total line intensity, it is approximately $10^{-4}$ per Gauss at wavelengths around 1 micron, which corresponds to only a few parts per billion of the disk spectral radiance. Previous measurements were demonstrated by Lin and coauthors in the early 2000s \citep{lin2000,lin2004}, but to achieve such a measurement over a useful field of view with sufficient resolution requires large aperture coronagraphic polarimetry.  

\begin{figure}
    \centering
    \includegraphics[width=0.485\linewidth]{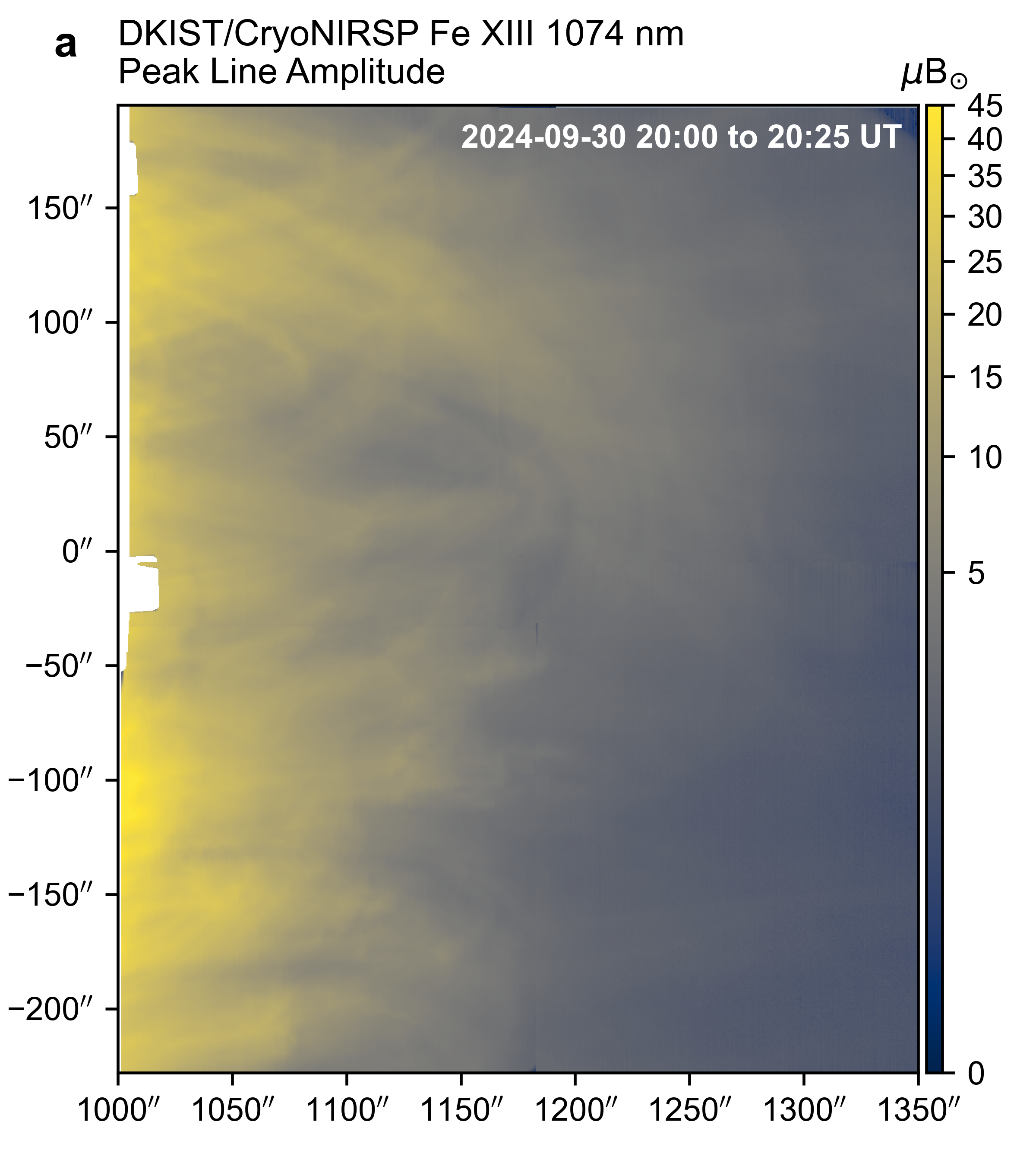}
    \includegraphics[width=0.485\linewidth]{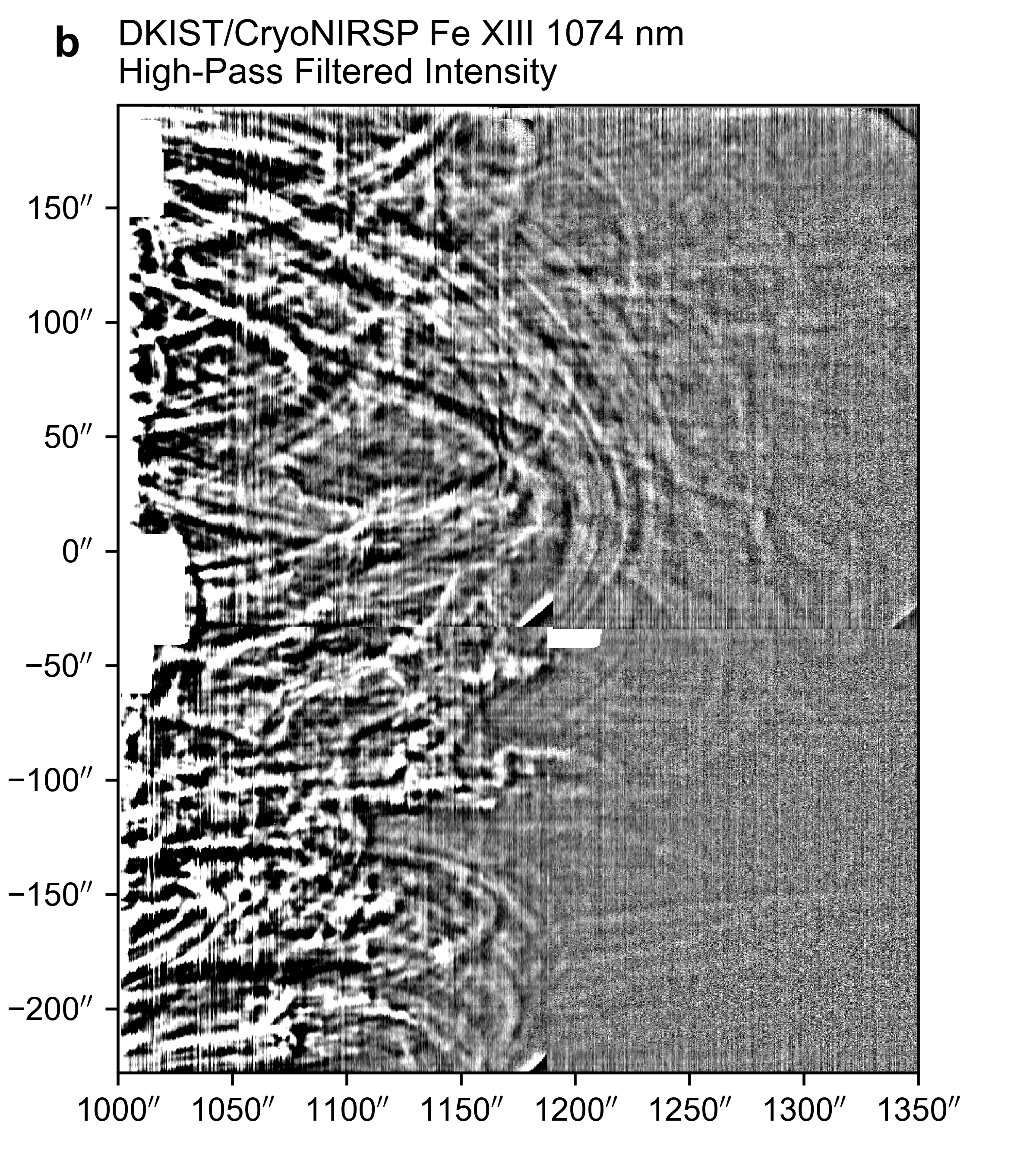} \\
    \includegraphics[width=0.485\linewidth]{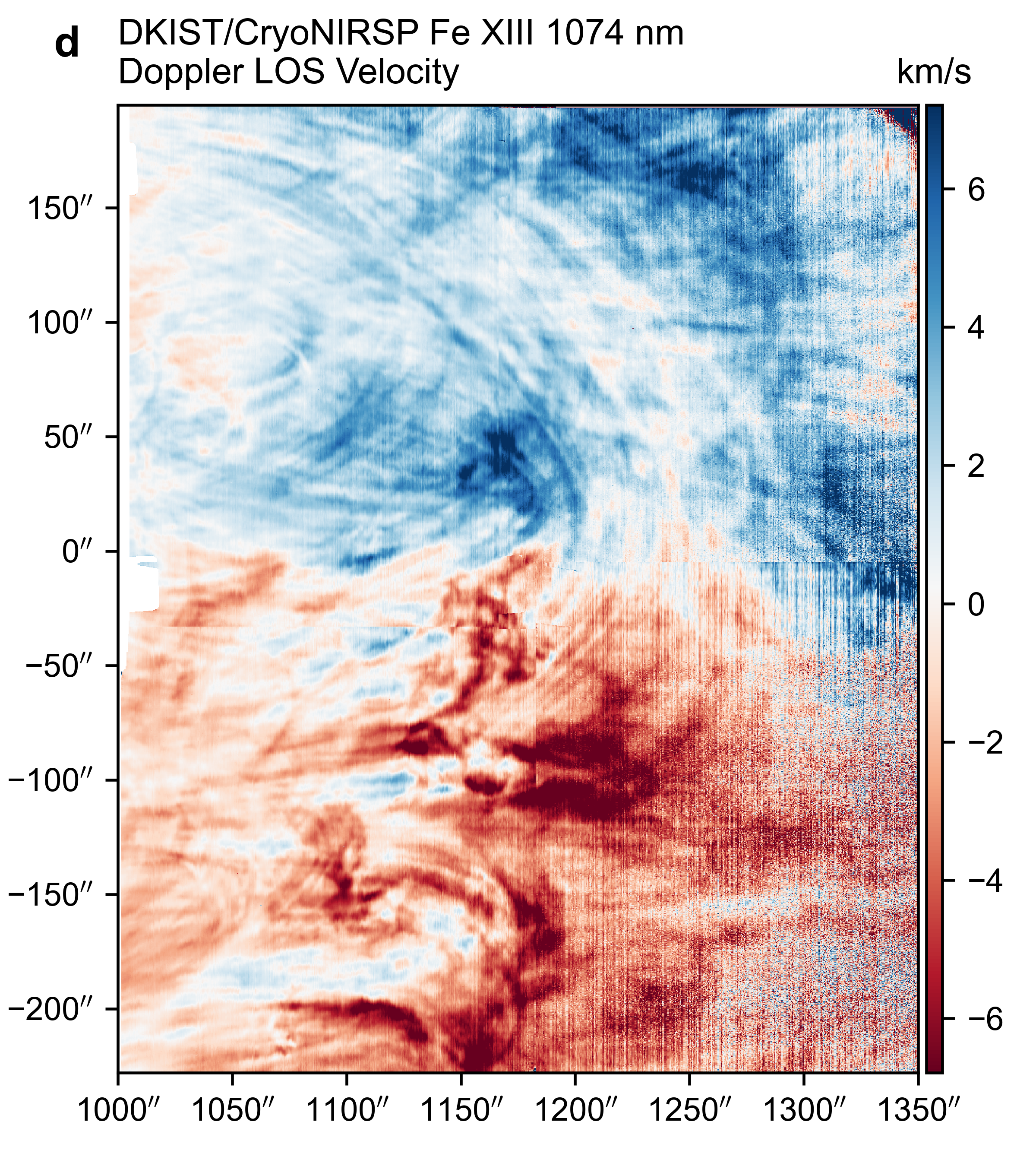}
    \includegraphics[width=0.485\linewidth]{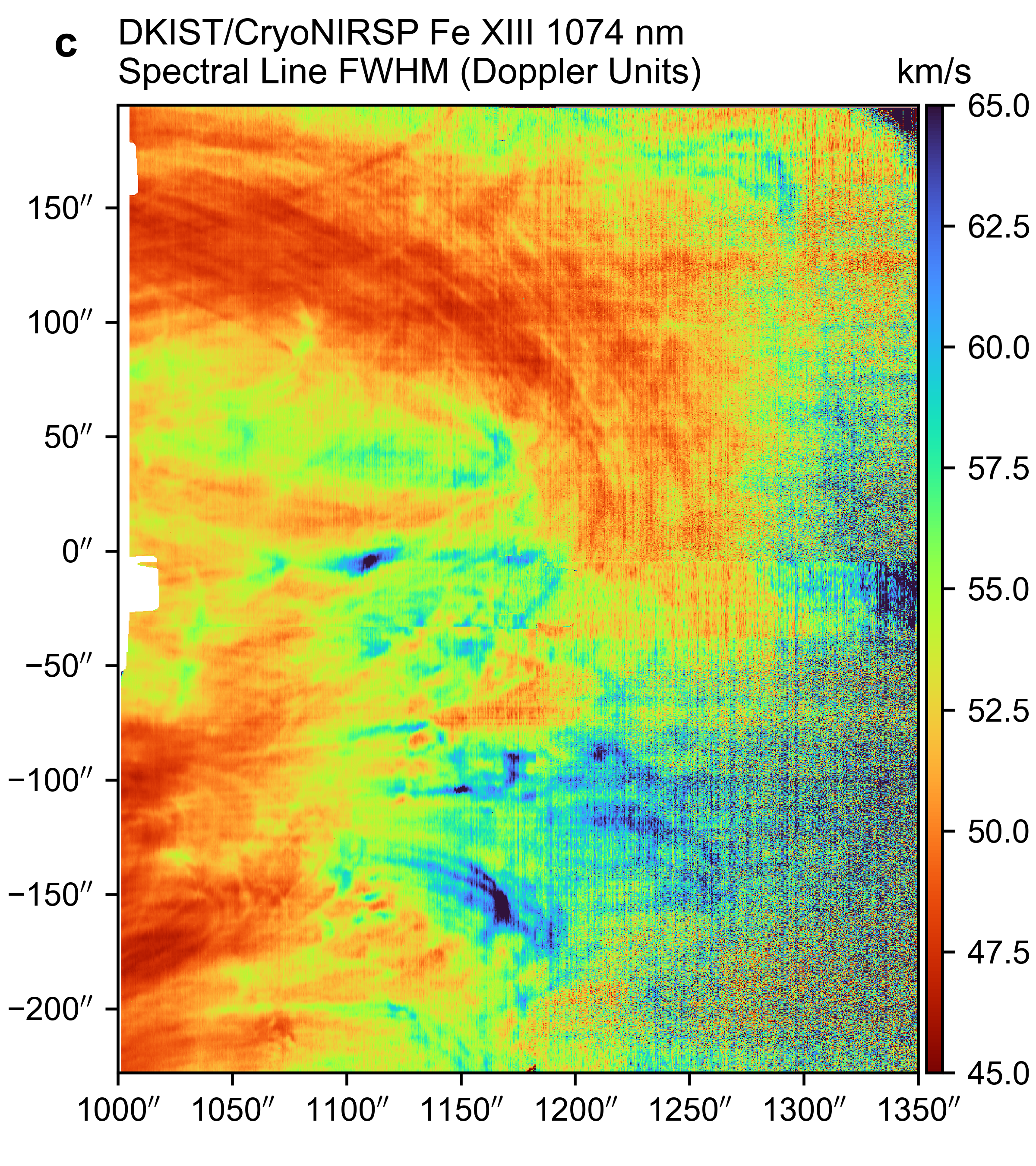}
    \caption{Large spectroscopic raster scans acquired by DKIST/CryoNIRSP on 30 September 2024 on the second day of a two-day program surrounding PSP Perihelion \#21. These maps result from coronal line fitting of the Fe XIII 1074 nm line across four separate raster scans. This is the largest contiguous field of view observation acquired by DKIST/CryoNIRSP to date. Not all available data are shown, and the analysis is preliminary (for display purposes only).}
    \label{fig:cn_2024}
\end{figure}

In June 2023, during a coordinated campaign around the PSP Encounter \#16, DKIST and its Cryo-NIRSP instrument \citep{fehlmann2023} obtained data that led to the first maps of the coronal magnetic field amplitude, shown in Figure~\ref{fig:stokesV} \citep{schad2024_science}.  These maps cover a field of view of approximately 400 x 200 arcseconds off-limb, acquired over three hours of observing time. Deep integrations and careful data calibration of the Fe XIII line at 1074 nm yielded circularly polarized signals concentrated within three coherent lobes of opposite polarity. The circularly polarized signals reach up to $\sim$25 parts per billion of the disk center spectral radiance;  when converted to an equivalent value in units of Gauss, this corresponds to 10--30 Gauss for the regions between 1 and 1.1 solar radii. 

Measuring the magnetic field in the corona is important for multiple reasons, including the impulsive release of energy in flares and CMEs, as well as its potential role for solar wind release and acceleration.   \cite{Wyper2022}, for example, suggests that intermittent interchange reconnection in active regions may produce a curtain of propagating and interacting torsional Alfv\`enic waves that could then develop into switchbacks within the heliosphere. For this reason, it is particularly valuable to advance coronal magnetometry with DKIST during PSP perihelia.   A more recent coordination occurred during PSP Perihelion \#21, with DKIST observing the west solar limb from Earth on Sept 29 and 30, 2024. Some high-quality large rasters of the active region that was likely connected to PSP during part of this period are shown in Figure~\ref{fig:cn_2024}. These maps extend beyond the edge of the SDO/AIA field of view and further provide Doppler shift and line width diagnostics. More extensive analysis of coronal densities and magnetic fields may be achievable with this multi-day dataset in the near future. 

\begin{table}
 \centering
 \caption{DKIST Cycle 1 and 2 data acquired during co-observing periods with PSP and/or Solar Orbiter.}\label{tbl:coord}
 {\small\begin{tabular}{lP{3cm}p{12mm}cP{3.5cm}}
    \midrule
    Date Range$^{1}$ & Event Description &   Exp. ID &  Instruments &  Select Data Notes \\
    \\
    \hline
    \textbf{2022} & & & & \\
    May 31 & PSP Perihelion \#12 &  EID\_1\_117 &  CryoNIRSP &    Data artifacts require manual processing \\ 
    June 2-3 & PSP disk passage after perihelion \#12 & \href{https://dkist.data.nso.edu/?proposalId=pid_1_118}{EID\_1\_118} & VBI + VISP &  Multiple publications already in print$^{2}$. See Figure~\ref{fig:vbi_2022} \\ 
    Oct 18-20 & Solar Orbiter RSW \#5 Long term AR SOOP & EID\_1\_122 & CryoNIRSP &  Data artifacts require manual processing  \\ 
    Oct 21,24 & Solar Orbiter RSW \#5 Long term AR SOOP & \href{https://dkist.data.nso.edu/?proposalId=pid_1_123}{EID\_1\_123} &  VBI + VISP & Hinode/SP coverage as well. \\ 
    \hline
    \textbf{2023} & & & & \\
    Mar 16-17 & PSP Perihelion \#15 & EID\_2\_109 & CryoNIRSP & Data artifacts require manual processing \\ 
    June 22 & PSP Perihelion \#16 & \href{https://dkist.data.nso.edu/?proposalId=pid_2_73}{EID\_2\_73} \newline \href{https://dkist.data.nso.edu/?proposalId=pid_2_6}{EID\_2\_6} & CryoNIRSP & \\ 
    Oct 6-7,10 & Solar Orbiter RSW \#11  AR long term SOOP & \href{https://dkist.data.nso.edu/?proposalId=pid_2_112}{EID\_2\_112} & CryoNIRSP &   \\ 
    Oct 6-7,10 & Solar Orbiter RSW \#11  Polar Field SOOP & \href{https://dkist.data.nso.edu/?proposalId=pid_2_113}{EID\_2\_113}  & VBI+VISP &   \\ 
    Oct 12-18 & Solar Orbiter RSW \#11 AR long term  SOOP  & \href{https://dkist.data.nso.edu/?proposalId=pid_2_114}{EID\_2\_114}  & VBI+VISP &  See Figure~\ref{fig:visp_2023} \\ 
    \hline
    {\textbf{2024}} & & & & \\
    Mar 23  & Solar Orbiter Prominence SOOP & \href{https://dkist.data.nso.edu/?proposalId=pid_2_119}{EID\_2\_119}  & CryoNIRSP &  \\ 
    Mar 29; Apr 8-9 &  Solar Eclipse & \href{https://dkist.data.nso.edu/?proposalId=pid_2_119}{EID\_2\_119}  & CryoNIRSP & Multiple datasets available including spectroscopy and polarimetry \\ 
    Sep 29 - Oct 8  &  PSP Encounter \#21 and Solar Orbiter RSW \#16 & \href{https://dkist.data.nso.edu/?proposalId=pid_2_127}{EID\_2\_127}  & CryoNIRSP & See Figure~\ref{fig:cn_2024}. Time series spectroscopy also available in Fe XIII 1074 nm. \\ 
    Oct 1-2 &   Solar Orbiter RSW \#17 Filament/Prominence SOOP & \href{https://dkist.data.nso.edu/?proposalId=pid_2_124}{EID\_2\_127} & VBI+VISP+DL &  \\  
    Oct 9,15 & Solar Orbiter RSW \#17 AR long term SOOP &  \href{https://dkist.data.nso.edu/?proposalId=pid_2_127}{EID\_2\_126} & VBI+VISP+DL  &  \\ 
    \midrule
    \end{tabular}}
\flushleft
\tabnote{\textit{Notes}: [1] Approximate date ranges only; for precise timings see the \href{https://dkist.data.nso.edu/dashboard}{DKIST Data Center Archive}. [2] \cite{daSilvaSantos2023,2024Judge,Kuridze2024} }
\end{table}
\section{DKIST-PSP-SO Joint Observing: Considerations and Existing Efforts}

In the above sections, we have broadly discussed how DKIST fits into the overall picture of multi-vantage point solar astronomy, and we have highlighted a few joint science questions. That said, a large number of science cases motivate joint DKIST, PSP, and Solar Orbiter observations. To put these into practice requires careful consideration of the different strengths, priorities, and operation models of these facilities. Unlike the mission-class experiments of PSP and Solar Orbiter, DKIST is a national observatory-class facility constructed for the use of the broad solar community.  It is made available via merit-based proposal calls that are granted potential observing time via a telescope allocation committee. This one-of-a-kind facility must also balance technical and science operations on a scale unlike any previous solar telescope while also being impacted by the local weather conditions, post-storm impacts, and seeing quality. 

Since the inauguration of DKIST as a facility in 2022, there have been two cycles of proposal-led science operations \citep{ParraguezCarcamo2024, Rimmele2024}.  Both are part of the DKIST Operations Commissioning Phase \citep{Tritschler2022}, during which the full life cycle of DKIST service-mode experiments has been exercised. This occurred alongside continued efforts to characterize, improve, and install instrumentation, while also bringing online the DKIST Data Center. Given the complexities and challenges faced during the OCP, no official coordination of DKIST with other facilities was offered via regular proposals. Instead, a limited amount of time was allocated to help support and test initial co-observing efforts during particularly important windows of joint observing opportunities. 

Table~\ref{tbl:coord} provides a high-level overview of joint observing periods to date between DKIST, PSP, and Solar Orbiter. Many of these observations occur around PSP perihelia, during which a DKIST off-limb coronal campaign is executed in succession with an on-disk program. The DKIST observing strategy during these times was developed through collaborative discussions between the mission science teams, with a particular focus on creating publicly available DKIST datasets that may help advance multi-point view science while also providing data that would broadly address multiple science cases identified in the DKIST Critical Science Plan \citep{Rast2021}. As can be gleaned from the table notes, the DKIST datasets produced have progressively improved over the first two DKIST observing cycles. The most recent campaign (in October 2024) has produced a number of valuable datasets, including the large map of the western limb during PSP Encounter \#21 shown in Figure~\ref{fig:cn_2024}. 

\section{Summary}

The current decade is one of intersecting frontiers for ground- and space-based solar missions, especially as DKIST, PSP, and Solar Orbiter have now come to fruition. Although these missions have been realized independently, there is a natural symbiotic relationship between them created by the complex interconnected questions of the Sun-heliosphere system.   Together, these facilities are helping to address the critical transitional interfaces that fundamentally govern the energy flow. From the DKIST perspective, we have emphasized how advancements in large aperture ground-based solar observations are pushing to resolve the bulk of the magneto-convective energy transport through the lower atmosphere. This is a critical step for measuring the energy flow throughout the heliosphere. Furthermore, we have discussed how advancing the frontier of chromospheric and coronal polarimetry, especially during quadrature opportunities, will soon help us understand magnetic connectivity and impulsive energy release. Continued efforts to unify these missions through separate and coordinated observations are essential.

\end{document}